\documentclass[published]{JHEP3} 

\JHEP{00(2002)000}

\JHEPspecialurl{http://jhep.sissa.it/JOURNAL/JHEP3.tar.gz}

\usepackage{epsfig,multicol}
         
\newcommand\fverb{\setbox\pippobox=\hbox\bgroup\verb}
\newcommand\fverbdo{\egroup\medskip\noindent%
\fbox{\unhbox\pippobox}\ }
\newcommand\fverbit{\egroup\item[\fbox{\unhbox\pippobox}]}
\newbox\pippobox

\title{Integrable string and hydrodynamical type models and nonlocal brackets}

\author{V. D. Gershun\\
	O. I. Akhieser Institute for Theoretical Physics, NSC Kharkiv Institute
of Physics and Technology, Academy of Sciences of Ukraine
	E-mail: \email{gershun@kipt.kharkov.ua}}
\received{, 2005} 		
\revised{, 2005}
\accepted{, 2005}

\abstract{The closed string model in the background gravity field is
considered as a bi-Hamiltonian system in assumption that string
model is the integrable model for particular kind of the
background fields.
 The dual nonlocal Poisson brackets (PB), depending of the background fields
and of their derivatives, are obtained. The integrability
condition is formulated as the compatibility of the bi-Hamiltonity
condition and the Jacobi identity of the dual Poisson bracket. It
is shown that the dual brackets and dual Hamiltonians can be
obtained from the canonical PB and from the initial Hamiltonian by
imposing the second kind constraints on the initial dynamical
system, on the closed string model in the constant background
fields, as example. The hydrodynamical type equation was obtained. 
Two types of the nonlocal brackets are
introduced. Constant curvature and time-dependent metrics are
considered, as examples. It is shown, that the Jacobi identities
for the nonlocal brackets have particular solution for the
space-time coordinates, as matrix representation of the simple Lie
group.}


\begin{document} 

\section{Introduction}
 The bi-Hamiltonian approach to the integrable systems was initiated by Magri
\cite{Ger:Mag} for the investigation of the integrability of the KdV equation
and  was generalized by Das, Okubo \cite{Ger:Das}.
{\bf Definition 1.}  A finite dimensional dynamical system with
$2N$ degrees of freedom $x^{a}, a=1,...2N$ is integrable, if it is
described by the set of the $n$ integrals of motion
$F_{1},...,F_{n}$ in involution under some Poisson bracket (PB)
\[\{F_{i}, F_{k}\}_{PB} = 0.\]
The dynamical system is completely solvable, if $n=N$. Any of the
integral of motion (or any linear combination of them) can be
considered as the
Hamiltonian $H_{k} = F_{k}$ .
{\bf Definition 2.}  The bi-Hamiltonity condition \cite{Ger:Das} has following
form:
\begin{equation}\label{eq1}\dot x^{a}= \frac{dx^{a}}{dt}=
\{x^{a},H_{1}\}_{1}=...= \{x^{a},H_{N}\}_{N}.\end{equation}
 The hierarchy of new PB is arose in this
connection:
\[{\{ ,\}_{1}, \{ ,\}_{2},..\{ ,\}_{N}}.\]
 The hierarchy of new dynamical systems arises under the new time
coordinates $t_{k}$.
\begin{equation}\label{eq2}\frac{dx^{a}}{dt_{n+k}}=
\{x^{a},H_{n}\}_{k+1}= \{x^{a},H_{k}\}_{n+1}.\end{equation} The
new equations of motion describe the new dynamical systems, which
are dual to the original system, with the dual set of the
integrals of motion.\\ 
 There is another approach to the bi-Hamiltonian systems \cite{Ger:Mag}. Two PB
$\{ , \}_{1}$ and $\{ , \}_{2}$ are called compatible if any linear 
combination of these PB $c_{1}\{ , \}_{1}+c_{2}\{ , \}_{2}$ is PB.
It is possible to find two corresponding Hamiltonians $H_{1}$ and 
$H_{2}$ which are satisfy to bi-Hamiltonity condition.\\
We used first approach to the closed string models as the 
bi-Hamiltonity systems. Second approach was used to description of the
hydrodynamical type models.\\  
 We consider the dynamical systems with constraints.
In this case, first kind constraints are generators of the gauge
transformations and they are integrals of motion. First kind
constrains $F_{k}(x^{a})\approx 0$, $k=1,2...$ form the algebra of
constraints under some PB.
\[\{F_{i}, F_{k}\}_{PB}= C_{ik}^{l}F_{l}\approx 0.\]
 The structure functions $C^{l}_{ik}$ may be functions of the
phase space coordinates in general case. The second kind
constraints $f_{k}(x^{a}) \approx 0$ are the representations of
the first kind constraints algebra. The second kind constraints is
defined by the condition
\[\{f_{i}, f_{k}\} = C_{ik}\ne 0.\]
The reversible matrix $C_{ik}$ is not constraint and also it is a
function of phase space coordinates. The second kind constraints
take part in deformation of the $\{ , \}_{PB}$ to the Dirac
bracket $\{ , \}_{D}$. As rule, such deformation leads to
nonlinear and to nonlocal brackets.
 The bi-Hamiltonity condition leads to the dual PB that are nonlinear and
nonlocal brackets as a rule. We suppose, that the dual brackets
can be obtained from the initial canonical bracket under the
imposition of the second kind constraints. We have applied
\cite{Ger:3,Ger:4,Ger:5,Ger:6},\,\,\cite{Ger:7,Ger:8,Ger:9,Ger:10} 
bi-Hamiltonian approach to the
investigation of the integrability of the closed string model in
the arbitrary background gravity field and antisymmetric B-field.
The bi-Hamiltonity condition and the Jacobi identities for the
dual brackets were considered as the integrability condition for a
closed string model. They led to some restrictions on the
background fields.\\
 Nonlocal PB of a hydrodynamical type was obtained as the Dirac bracket
by Ferapontov \cite{Ger:Fer2} and by Maltsev \cite{Ger:Mal1}.\\
  The plan of the paper is the following. In the second section we briefly
considered papers about hydrodynamical type nonlocal brackets. 
In the third section we considered closed string model in the arbitrary 
background gravity field. We suppose that
this model is an integrable model for some configurations of the
background fields. The bi-Hamiltonity condition and the Jacobi
identities for the dual PB resulted in to the integrability
condition, which restrict the possible configurations of the
background fields. As examples we considered constant curvature
space and time-dependent metric space. In the fourth section we
considered closed string model in the constant background gravity field.
We obtained hydrodynamical type equation for the string model on the second
kind constraints as configuration subspace embedded in a phase space.

\section{Hydrodynamical type models}
 Mokhov and Ferapontov introduced the nonlocal PB
\cite{Ger:Mok1}. The Ferapontov nonlocal
PB (or hydrodynamical type nonlocal PB ) \cite{Ger:Fer1} is:
\begin{equation}\label{eq3}
\{u^{i}(x),u^{k}(y)\}=g^{ik}(u)\frac{\partial}{\partial x}
\delta (x-y)-g^{ij}\Gamma^{k}_{jl}u^{l}_{x}\delta (x-y)+
\sum_{s=1}^{L}\omega^{(s)i}_{j}(u(x))u^{j}_{x}\nu
(x-y)\omega^{(s)k}_{l}(u(y))u^{l}_{y}, \end{equation} 
where $\nu
(x-y)=sgn (x-y)=(\frac{d}{dx})^{-1}\delta (x-y)$, $u^{i}(x)$ are local
coordinates, $u_{x}^{i}(x)=\partial_{x}u^{i}(x), i=1...N$. The coefficients 
$g^{ik}(x),\Gamma^{k}_{jl}(x),\omega ^{(s)i}_{k}(x)$ are smooth functions 
of local coordinates.
 This nonlocal PB is satisfy the Jacoby identity if and only if
$g^{ik}(u)$ is the pseudo-Riemannian metric without torsion and also the
coefficients satisfy the following relations:\\
1. $\Gamma ^{k}_{jl}(u)$ is the Levi-Civita connection,\\
2. $g^{ik}(u)\omega ^{(s)j}_{k}(u)=g^{jk}(u)\omega ^{(s)i}_{k}(u),$\\
3. $\nabla _{k}\omega^{(s)i}_{l}(u)=\nabla _{l}\omega^{(s)i}_{k}$, where
$\nabla _{k}$ is the covariant differential,\\
4. $R^{ij}_{kl}(u)=\sum_{s=1}^{L}[\omega^{(s)i}_{l}\omega^{(s)j}_{k}-
\omega^{(s)j}_{l}\omega^{(s)i}_{k}]$, where $R^{ij}_{kl}$ is Riemannian
curvature tensor of the metric $g^{ik}$,\\
5. $\omega^{(s)i}_{k}\omega^{(t)k}_{l}=\omega^{(t)i}_{k}\omega^{(s)i}_{k}.$\\
 This nonlocal PB corresponds to an N-dimensional surface with flat normal 
bundle embedded in a pseudo-Euclidean space $E^{N+L}$ \cite{Ger:Fer2}. There 
metric $g^{ik}$ is the first fundamental form,
$\omega ^{(s)i}_{k}$ is Weingarten operator of this embedded surface, which
is define the second fundamental form. The relations 2)-4) are the
Gauss-Peterson-Codazzi equations. The relations 5) are correspond to the
Ricci equations for this embedded surface. \\
 Dubrovin and Novikov have considered the local
dual PB of the similar type \cite{Ger:Dub} in the
application to the Hamiltonian hydrodynamical models. Dubrovin-Novikov PB 
( or the hydrodynamical type local PB) can be obtained from the nonlocal PB   
(\ref{eq3}) under condition $\omega ^{(s)i}_{k}=0$.
 The Jacobi identity for this PB is satisfied if
$g_{ik}$ is the Riemann metric without torsion, the curvature
tensor is equal to zero. The metric tensor is constant, locally.\\
 It need to consider the linear combination of the local and the nonlocal
Poisson brackets to obtain the hydrodynamical type equations \cite{Ger:Mal}.
 There we consider Mokhov, Ferapontov, nonlocal PB \cite{Ger:Mok1} for the 
metric space of constant Riemannian curvature K, as example:
\begin{eqnarray}\label{eq4}
\{u^{i}(x),u^{k}(y)\}=c_{1}\eta ^{ik}\frac{d}{dx}\delta (x-y)+
c_{2}(\frac{\partial h^{k}}{\partial u^{i}}+\frac{\partial h^{i}}
{\partial u^{k}}-Ku^{i}u^{k})\frac {d}{dx}\delta (x-y)\nonumber\\
+(\frac{{\partial}^{2}h^{k}}{\partial u^{i}\partial u^{l}}-
K\delta ^{i}_{l}u^{k})u^{l}_{x}\delta (x-y)+Ku^{i}_{x}\nu (x-y)u^{k}_{y}.
\end{eqnarray}
The canonical form of PB (\ref{eq4}) was first presented by Pavlov 
\cite{Ger:Pav1}.
 The Jacobi identity is satisfied on the following relations:
\[
\frac {{\partial}^{2}h^{i}}{\partial u^{k}\partial u^{n}}\frac {{\partial}^{2}
h^{j}}{\partial u^{n}\partial u^{l}}= \frac {{\partial}^{2}h^{j}}
{\partial u^{k}
\partial u^{n}}\frac {{\partial}^{2}h^{i}}{\partial u^{n}\partial u^{l}},
\]
\[(\frac{\partial h^{n}}{\partial u^{i}}+
\frac{\partial h^{i}}
{\partial u^{n}}-
Ku^{i}u^{n})\frac{{\partial}^{2}h^{k}}{\partial u^{j}\partial u^{n}}=
\{i \longleftrightarrow j \}.\]
 First of this equations is the WDVV \cite{Ger:Wit,Ger:Ver} consistence 
local condition.
The system of hydrodynamical type is a bi-Hamiltonian system with the PB
$\{ , \}_{FM}$ and $\{ , \}_{ND}$  if:
\[\dot u^{i}(x)=\{u^{i}(x), H_{1}\}_{FM}=\{u^{i}(x), H_{2}\}_{ND}.\]
Where Hamiltonians $H_{1}$ and $H_{2}$ are following:
\[H_{1}=\frac{1}{2}\int u^{i}(x)u^{i}(x)dx,\,\,H_{2}=\int [h^{i}(u(x))u^{i}(x)-
\frac{K}{8}u^{i}u^{i}u^{k}u^{k}]dx.\]
\section{ Closed string in the background fields.}
 The string model in the background gravity field is described by the
system of the equations:
\[\ddot x^{a}-x^{''a}+\Gamma^{a}_{bc}(x)(\dot x^{b}\dot
x^{c}-x^{'b}x^{'c})=0,\,\, g_{ab}(x)(\dot x^{a}\dot
x^{b}+x^{'a}x^{'b})=0,\,\, g_{ab}(x)\dot x^{a}x^{'b}=0,\nonumber\\
\]
where $\dot x^{a}=\frac{dx^{a}}{d \tau}$, $x^{'a}=\frac{dx^{a}}{d
\sigma}$.
 We will
consider the Hamiltonian formalism. The closed string in the
background gravity field is described by first kind constraints in
the Hamiltonian formalism:
\begin{equation}\label{eq5}h_{1}=\frac{1}{2}g^{ab}(x)p_{a}p_{b}
+\frac{1}{2}g_{ab}(x)x^{'a}x^{'b}\approx 0,\,\,
h_{2}=p_{a}x^{'a}\approx 0,\end{equation} where $a,b =0,1,...D-1$,
$x^{a}(\tau,\sigma), p_{a}(\tau,\sigma)$ are the periodical
functions on $\sigma$ with the period on $\pi$. The original PB
are the canonical  PB:
\[\{x^{a}(\sigma),p_{b}({\sigma}')\}_{1}=\delta _{b}^{a}
\delta (\sigma -{\sigma}'), \{x^{a}(\sigma),x^{b}({\sigma}'
\}_{1}=\{p_{a}(\sigma),p_{b}({\sigma}') \}_{1}=0.\] The
Hamiltonian equations of motion of the closed string, in the
arbitrary background gravity field under the Hamiltonian
$H_{1}=\int\limits_{0}^{\pi}h_{1}d\sigma$ and PB $\{,\}_{1}$, are
\[\dot x^{a}= g^{ab}p_{b},\,\,\, \dot p_{a}= g_{ab}x^{''b}-
\frac{1}{2}\frac{\partial g^{bc}}{\partial x^{a}}p_{b}p_{c}
-\frac{1}{2}\frac{\partial g_{bc}}{\partial x^{a}}x'^{b}x'^{c}+
\frac{\partial g_{ac}}{\partial x^{b}}x'^{b}x'^{c}.\] The dual PB are 
obtained from the bi-Hamiltonity condition
\begin{equation}\label{eq6}\dot
x^{a}=\{x^{a},\int\limits_{0}^{\pi}h_{1}({\sigma}')d{\sigma}'
\}_{1}= \{x^{a},\int\limits_{0}^{\pi}h_{2}({\sigma}')d{\sigma}'
\}_{2},\end{equation}
\[\dot
p_{a}=\{p_{a},\int\limits_{0}^{\pi}h_{1}({\sigma}')d{\sigma}'\}_{1}=
\{p_{a},\int\limits_{0}^{\pi}h_{2}({\sigma}')d{\sigma}'\}_{2}.\]
They have the following form:

{\bf Proposition 1.}
\[\{A(\sigma ),B({\sigma}' )\}_{2}=
\frac{\partial A}{\partial x^{a}}\frac{\partial B}{\partial
x^{b}}[[\omega^{ab} (\sigma )+\omega^{ab}({\sigma}' )]\nu
({\sigma}' -\sigma )+ [\Phi^{ab}(\sigma )+ \Phi^{ab}({\sigma}'
)]\frac{\partial}{\partial {\sigma}' }\delta ({\sigma}'
 -\sigma )\]
\[+[\Omega^{ab}(\sigma )+\Omega^{ab}({\sigma}' )]\delta ({\sigma}' -\sigma )]+
\frac{\partial A}{\partial p_{a}}\frac{\partial B}{\partial
p_{b}}[[\omega_{ab} (\sigma )+\omega_{ab}({\sigma}' )]\nu
({\sigma}' -\sigma )+\]
\[+[\Phi_{ab}(\sigma )+\Phi_{ab}({\sigma}' )]\frac{\partial}
{\partial {\sigma}' } \delta ({\sigma}' -\sigma )+
[\Omega_{ab}(\sigma )+\Omega_{ab}({\sigma}' )] \delta ({\sigma}'
-\sigma )]+\]
\[ +[\frac{\partial A}{\partial x^{a}}\frac{\partial B}{\partial
p_{b}}+\frac{\partial A}{\partial p_{b}}\frac{\partial B}{\partial
x^{a}}][[\omega_{b}^{a}(\sigma )+\omega_{b}^{a}({\sigma}' )]\nu
({\sigma}' -\sigma )+ [\Phi_{b}^{a}(\sigma
)+\Phi_{b}^{a}({\sigma}' )]\frac{\partial}{\partial {\sigma}'
}\delta ({\sigma}' -\sigma )]\]
\[ + [\frac{\partial A}{\partial x^{a}}\frac{\partial B}{\partial
p_{b}}-\frac{\partial A}{\partial p_{b}}\frac{\partial B}{\partial
x^{a}}] [\Omega _{b}^{a}(\sigma)+ \Omega_{b}^{a}({\sigma}'
)]\delta ({\sigma}' - \sigma )\] The arbitrary functions $A, B,
\omega, \Phi, \Omega$ are the functions of the $x^{a}(\sigma ),
p_{a}(\sigma )$.  The functions $\omega^{ab}, \omega_{ab}$,
$\Phi^{ab},\Phi_{ab}$ are the symmetric functions on $a, b$ and
$\Omega^{ab}, \Omega_{ab}$ are the antisymmetric functions to
satisfy the condition $\{A, B\}_{2}=-\{B,A\}_{2}$.
The equations of motion under the
Hamiltonian $H_{2}=\int\limits_{0}^{\pi}
h_{2}({\sigma}')d{\sigma}'$ and PB $\{, \}_{2}$ are
\[ \dot
x^{a}=-\omega_{b}^{a}x^{b}+2\omega^{ab}p_{b}+2\Phi^{ab}p_{b}^{''}
-2\Phi_{b}^{a}x^{''b}+2\Omega_{b}^{a}x^{'b}-2\Omega^{ab}p_{b}^{'}+
\]
\[+ \int\limits_{0}^{\pi}d{\sigma}' [\omega_{b}^{a}x^{'a}+\frac{d
\omega^{ab}}{d{\sigma}'}p_{b}]\nu ({\sigma}' -\sigma ) +\frac{d
\Phi^{ab}}{d \sigma}p_{b}^{'} -\frac{d \Phi_{b}^{a}}{d
\sigma}x^{'b},\]
\[ \dot
p_{a}=-\omega_{ab}x^{b}-2\Phi_{ab}x^{''b}+2\Omega_{ab}x^{'b}+
2\omega_{a}^{b}p_{b}+2\Phi_{a}^{b}p_{b}^{''}+2\Omega_{a}^{b}p_{b}^{'}+\]
\[+ \int\limits_{0}^{\pi}d{\sigma}' [\omega_{ab}x^{'b}+\frac{d
\omega^{b}_{a}}{d{\sigma}'}p_{b}]\nu ({\sigma}' -\sigma ) -\frac{d
\Phi_{ab}}{d \sigma}x^{'b}+ \frac{d \Phi_{a}^{b}}{d
\sigma}p_{b}^{'}.\]
 The bi-Hamiltonity condition (\ref{eq6}) is led to the two constraints
\[
-\omega_{b}^{a}x^{b}+2\omega^{ab}p_{b}+2\Phi^{ab}p_{b}^{''}
-2\Phi_{b}^{a}x^{''b}+2\Omega_{b}^{a}x^{'b}-2\Omega^{ab}p_{b}^{'}+
\]
\[\int\limits_{0}^{\pi}d{\sigma}' [\omega_{b}^{a}x^{'a}+\frac{d
\omega^{ab}}{d {\sigma}'}p_{b}]\nu ({\sigma}' -\sigma ) +\frac{d
\Phi^{ab}}{d \sigma}p_{b}^{'}- \frac{d \Phi_{b}^{a}}{d
\sigma}x^{'b}=g^{ab}p_{b},\]

\[-\omega_{ab}x^{b}-2\Phi_{ab}x^{''b}+2\Omega_{ab}x^{'b}+
2\omega_{a}^{b}p_{b}+2\Phi_{a}^{b}p_{b}^{''}+2\Omega_{a}^{b}p_{b}^{'}+\]
\[ + \int\limits_{0}^{\pi}d{\sigma}' [\omega_{ab}x^{'b}+
\frac{d \omega^{b}_{a}}{d{\sigma}'}p_{b}]\nu ({\sigma}' -\sigma )
-\frac{d \Phi_{ab}}{d \sigma}x^{'b}+ \frac{d\Phi_{a}^{b}}{d
\sigma}p_{b}^{'}= \]
\[+g_{ab}x^{''b}- \frac{1}{2}\frac{\partial
g^{bc}}{\partial x^{a}}p_{b}p_{c} -\frac{1}{2}\frac{\partial
g_{bc}} {\partial x^{a}}x^{'b}x^{'c}+ \frac{\partial
g_{ac}}{\partial x^{b}}x^{'b}x^{'c}.\]

In really, there is the list of the constraints depending on the
possible choice of the unknown functions $\omega$, $\Omega$, $\Phi$.
 In the general case, there are both
the first kind constraints and the second kind constraints. Also
it is possible to solve the constraints equations as the equations
for the definition of the functions $\omega$,$\Phi$,$\Omega$. We
considered the latter possibility and we obtained the following
consistent solution of the bi-Hamiltonity condition:
\begin{eqnarray}\Phi^{ab}=0,\,\,\Omega^{ab}=0,\,\,\Phi_{b}^{a}=0,\,\,
\Omega ^{a}_{b}=0,\,\,
\frac{\partial\omega^{ab}}{\partial x^{c}}x^{c}+2\omega^{ab}=g^{ab},\nonumber\\
\omega_{ab}=\frac{1}{2}\frac{\partial^{2} \omega^{cd}}{\partial
x^{a}\partial x^{b}}p_{c}p_{d},\,\,\omega_{b}^{a}=-\frac{\partial
\omega^{ac}}{\partial
x^{b}}p_{c},\nonumber \\
\Phi_{ab}=-\frac{1}{2}g_{ab},
\Omega_{ab}=\frac{1}{2}(\frac{\partial \Phi_{bc}}{\partial x^{a}}-
\frac{\partial \Phi_{ac}}{\partial x^{b}})x^{'c},\,\,
\frac{\partial\omega^{ab}} {\partial p_{c}}=0.
\nonumber\end{eqnarray}\\
{\bf Remark 1.} In distinct from the PB of the hydrodynamical
type, we need to introduce the separate PB for the coordinates of
the Minkowski space and for the momenta because, the gravity field
is not depend of the momenta. Although, this difference is
vanished under the such constraint as $f(x^{a},p_{a})\approx 0$.\\
Consequently, the dual PB for the phase space coordinates are
\[\{x^{a}(\sigma
),x^{b}({\sigma}')\}_{2}= [\omega^{ab}(\sigma
)+\omega^{ab}({\sigma}' )]\nu ({\sigma}' - \sigma ),\]
\[ \{p_{a}(\sigma ),p_{b}({\sigma}' )\}_{2}=
[\frac{\partial^{2} \omega_{cd}(\sigma )}{\partial x^{a}\partial
x^{b}}p_{c}p_{d}+\frac{\partial^{2} \omega_{cd}({\sigma}'
)}{\partial x^{a}\partial x^{b}}p_{c}p_{d}]\nu ({\sigma}' - \sigma
)- \]
\[-\frac{1}{2}[g_{ab}(\sigma)+g_{ab}({\sigma}')]\frac{\partial}{\partial
{\sigma}' }\delta ({\sigma}' - \sigma )+ [\frac{\partial
g_{ac}}{\partial x^{b}}-\frac{\partial g_{bc}}{\partial
x^{a}}]x^{'c}(\sigma)\delta ({\sigma}' - \sigma)\]
\[ \{x^{a}(\sigma ),p_{b}({\sigma}' )\}_{2}= -
[\frac{\partial \omega^{ac}(\sigma )}{\partial
x^{b}}p_{c}+\frac{\partial \omega^{ac}({\sigma}')}{\partial
x^{b}}p_{c} ]\nu ({\sigma}' - \sigma ),\]
\begin{equation}\label{eq7}\{p_{a}(\sigma ),x^{b}({\sigma}'
)\}_{2}= - [\frac{\partial \omega^{bc}(\sigma)}{\partial
x^{a}}p_{c}+\frac{\partial \omega^{bc}({\sigma}' )}{\partial x^{c}
}p_{c}]\nu ({\sigma}' -\sigma ).\end{equation}
 The function
$\omega^{ab}(x)$ is satisfied on the equation:
\begin{equation}\label{eq8}\frac{\partial \omega^{ab}}{\partial
x^{c}}x^{c}+2\omega^{ab} =g^{ab}.\end{equation} The Jacobi
identities for the PB $\{,\}_{2}$ are led to the nonlocal
consistence conditions on the unknown function
$\omega^{ab}(\sigma)$. We can calculate unknown metric tensor
$g^{ab}(\sigma)$ by substitution of the solution of the
consistence condition for $\omega^{ab}$ to the equation
(\ref{eq8}).
  The Jacobi identity
\begin{equation}\label{eq9}
\{x^{a}(\sigma),x^{b}({\sigma}')\}x^{c}({\sigma}'')\}_{J}\equiv\end{equation}
\[\{x^{a}(\sigma),x^{b}({\sigma}')\}x^{c}({\sigma}'')\}
+\{x^{c}({\sigma}''),x^{a}(\sigma)\}x^{b}({\sigma}')\}+\{x^{b}({\sigma}'),
x^{c}({\sigma}'')\}x^{a}(\sigma\})=0\] is led to the following
nonlocal analogy of the WDVV \cite{Ger:Wit, Ger:Ver}
consistence condition:
\[[\frac{\partial \omega ^{ab}(\sigma)}{\partial
x^{d}}[\omega^{dc}(\sigma)+\omega^{dc}({\sigma}'')]-
\frac{\partial \omega^{ac}(\sigma)}{\partial x^{d}}[\omega
^{db}(\sigma)+\omega^{db}({\sigma}')]]\nu ({\sigma}'-\sigma)\nu
({\sigma}''- \sigma)+\]
\[ [\frac{\partial \omega ^{cb}({\sigma}')}{\partial
x^{d}}[\omega^{da}({\sigma}')+\omega^{da}(\sigma)]-
 \frac{\partial
\omega^{ab}({\sigma}')}{\partial x^{d}}[\omega
^{dc}({\sigma}')+\omega^{dc}({\sigma}'')]]\nu
(\sigma-{\sigma}')\nu ({\sigma}'' - {\sigma}')+\nonumber\]
\begin{equation}\label{eq10}[\frac{\partial \omega^{ac}({\sigma}'')}
{\partial x^{d}} [\omega^{db}({\sigma}'')+\omega^{db}({\sigma}')]-
\frac{\partial \omega^{cb} ({\sigma}'')}{\partial
x^{d}}[\omega^{da}({\sigma}'')+ \omega^{da}(\sigma)]] \nu
(\sigma-{\sigma}'')\nu ({\sigma}'-{\sigma}'')=0.\end{equation}

  This equation has the particular solution of the following form:
\[\frac{\partial\omega^{ab}(\sigma)}
{\partial x^{d}}
[\omega^{dc}(\sigma)+\omega^{dc}({\sigma}'')]-\frac{\partial
\omega^{ac} (\sigma)}{\partial
x^{d}}[\omega^{db}(\sigma)+\omega^{db}({\sigma}')]= \nonumber \]
\[ [T^{b},T^{c}]T^{a}]f(\sigma,{\sigma}',{\sigma}'')\nu({\sigma}''-
\sigma)\nu ({\sigma}'-\sigma),\] where $T^{a}, a = 0,1,...D-1$ is
the matrix representation of the simple Lie algebra and
$f(\sigma,{\sigma}',{\sigma}'')$ is arbitrary function. The Jacobi
identity is satisfied on the Jacobi identity of the simple Lie
algebra in this case:
\[([T^{a},T^{b}]T^{c}]+[T^{c},T^{a}]T^{b}]+
[T^{b},T^{c}]T^{a}])f(\sigma,{\sigma}',{\sigma}'')=0\] and we used
the relation ${\nu}^{2}({\sigma}'-\sigma)=1$.
 The local solution of the Jacobi identities leads to the constant metric
tensor. The rest Jacobi identities are cumbrous and we do not
reduce this expressions here.
 The symmetric factor of $\sigma,{\sigma}'$ of the antisymmetric functions
$\nu({\sigma}'-\sigma)$,$\frac{\partial}{\partial
\sigma}\delta(\sigma-{\sigma}')$ in the right side of the PB can
be both sum of the functions of $\sigma$ and ${\sigma}'$, and
production of them.
 Last possibility can be used in the vielbein formalism.\\
{\bf Proposition 2.}  The bi-Hamiltonity condition can be solved in
the terms PB $\{,\}_{2}$, which have the following form:
\[\{x^{a}(\sigma),x^{b}({\sigma}')\}_{2}=
e^{a}_{\mu}(\sigma)e^{b}_{\mu}({\sigma}')\nu ({\sigma}'-\sigma),\]
\[\{x^{a}(\sigma),p_{b}({\sigma}'\}_{2}=-e^{a}_{\mu}
(\sigma)\frac{\partial e^{c}_{\mu}({\sigma}')}{\partial
x^{b}}p_{c}({\sigma}') \nu ({\sigma}'-\sigma),\]
\[\{p_{a}(\sigma),p_{b}({\sigma}')\}_{2}=\frac{\partial
e^{c}_{\mu}(\sigma)}{\partial x^{a}}p_{c}(\sigma)\frac{\partial
e^{d}_{\mu} ({\sigma}')}{\partial x^{b}}p_{d}({\sigma}')\nu
({\sigma}'-\sigma)-
e_{a}^{\mu}(\sigma)e_{b}^{\mu}({\sigma}')\frac{\partial }{\partial
{\sigma}'} \delta ({\sigma}'-\sigma)+\]
\begin{equation}\label{eq11}+[\frac{\partial e_{a}^{\mu}}
{\partial x^{c}}e_{b}^{\mu} -\frac{\partial e_{b}^{\mu}}{\partial
x^{c}}e_{a}^{\mu}-\frac {\partial e_{c}^{\mu}}{\partial
x^{a}}e_{b}^{\mu}+ \frac{\partial e_{c}^{\mu}}{\partial
x^{b}}e_{a}^{\mu}] x^{'c}(\sigma)\delta
({\sigma}'-\sigma),\end{equation} where veilbein $e^{a}_{\mu}$ is
satisfied on the additional conditions:
\[g^{ab}= \eta ^{\mu\nu}e^{a}_{\mu}e^{b}_{\nu},\,\,
g_{ab}=\eta_{\mu\nu}e_{a}^{\mu}e_{b}^{\nu}\]
and $\eta^{\mu\nu}$ is the metric tensor of the flat space.

The particular solution of the Jacobi identity is
\[\frac{\partial e^{a}_{\mu}(\sigma)}
{\partial x^{d}}e^{b}_{\mu}({\sigma}')
e^{d}_{\nu}(\sigma)e^{c}_{\nu}({\sigma}'')-\frac{\partial
e^{a}_{\mu} (\sigma)}{\partial
x^{d}}e^{c}_{\mu}({\sigma}'')e^{d}_{\nu}(\sigma)e^{b}_
{\nu}({\sigma}')=\]
\[[T^{b},T^{c}]T^{a}]f(\sigma,{\sigma}',{\sigma}'')\nu ({\sigma}''-\sigma)
\nu ({\sigma}'-\sigma).\]
As example let me consider the the constant curvature space.\\
{\bf Example 1.} The constant curvature space is described by the
metric tensor $g_{ab}(x(\sigma))$ and by it inverse tensor
$g^{-1}_{ab}$:
\[g_{ab}=\eta_{ab}+\frac{kx_{a}x_{b}}{1-kx^{2}}, \,\,
g^{ab}\equiv g^{-1}_{ab}=\eta_{ab}-kx_{a}x_{b}.\]\\
{\bf Proposition 3.} Dual (PB)$\{,\}_{2}$ are:
\[\{x_{a}(\sigma),x_{b}({\sigma}')\}=[\eta _{ab}-kx_{a}
(\sigma)x_{b}({\sigma}')] \nu ({\sigma}'-\sigma),\]
\[\{x_{a}(\sigma),p_{b}({\sigma}')\}=
kx_{a}(\sigma)p_{b}({\sigma}')\nu ({\sigma}'-\sigma),\]
\[\{p_{a}(\sigma),p_{b}({\sigma}')\}=-kp_{a}(\sigma)p_{b}
({\sigma}')\nu ({\sigma}'- \sigma)\]
\begin{equation}\label{eq12}-\frac{1}{2}[2\eta _{ab}+
\frac{kx_{a}x_{b}}{1-kx^{2}}(\sigma)+
\frac{kx_{a}x_{b}}{1-kx^{2}}({\sigma}')]\frac{\partial}{\partial
{\sigma}'} \delta ({\sigma}'-\sigma)+ \frac{x_{a}x'_{b}-
x_{b}x'_{a}}{2(1-kx^{2})}\delta ({\sigma}'-\sigma).\end{equation}

The Jacobi identity (\ref{eq9}) is led to the equation
\[[\eta_{ab}x_{c}({\sigma}'')-\eta_{ac}x_{b}({\sigma}')]
\nu ({\sigma}'-\sigma)\nu
(\sigma-{\sigma}'')+[\eta_{bc}x_{a}(\sigma)-
\eta_{ba}x_{c}({\sigma}'')] \nu (\sigma-{\sigma}')\nu
({\sigma}'-{\sigma}'')+\nonumber \]
\[[\eta_{ca}x_{b}({\sigma}')
-\eta_{cb}x_{a}(\sigma)] \nu ({\sigma}'-{\sigma}'')\nu
({\sigma}''-\sigma)=0.\nonumber \] The particular solution of this
equation is:
\begin{equation}\label{eq13}\eta_{ab}x_{c}({\sigma}'')-\eta_{ac}x_{b}
({\sigma}')= [T_{b},T_{c}]T_{a}]f(\sigma,{\sigma}',{\sigma}'')\nu
({\sigma}''-\sigma)\nu ({\sigma}'-\sigma).\end{equation}
 Consequently, the space-time coordinate $x_{a}(\sigma)$ is the matrix
representation of the simple Lie algebra.
  The Jacobi identity
 $\{x_{a}(\sigma),x_{b}({\sigma}')\}p_{c}({\sigma}'')\}_{J}$ is led to
the equation
\begin{equation}\label{eq14}k\eta_{ab}p_{c}({\sigma}'')\nu
({\sigma}'-\sigma)[\nu ({\sigma}''-\sigma)+\nu
({\sigma}''-{\sigma}')]=0.
\end{equation}
 These results can be obtained from the veilbein formalism under the following
ansatz for the veilbein of the constant curvature space:
\[e^{a(s)}_{\mu}=n_{\mu}(m^{(s)}_{1}n^{a}+
\sqrt{-k}m^{(s)}_{2}x^{a}), e_{a}^{\mu
(s)}=n^{\mu}g_{ab}(m^{(s)}_{1}n^{b}+\sqrt{-k}m^{(s)}_{2}x^{b}),
\]
where $n^{2}_{\mu}=1,\,\,m^{(s)}_{1}m^{(s)}_{1}=1,\,\,
m^{(s)}_{2}m^{(s)}_{2}=1,\,\,m^{(s)}_{1}m^{(s)}_{2}=0,\,\,n^{a}n^{b}=
\delta^{ab}$ and $(s)$ is number of the solution of the equations
\[e^{a}_{\mu}e^{b}_{\mu}=g^{ab},\,\,e_{a}^{\mu}e_{b}^{\mu}=g_{ab},\,\,
e^{a}_{\mu}e^{\mu}_{b}=\delta ^{a}_{b}.\]
  The following example is time-dependent metric space.\\
{\bf Example 2.} The time-dependent metric in the light-cone
variables has form:
\label{eq15}
\begin{equation}ds^{2}=g_{ik}(x^{+})dx^{i}dx^{k}+g_{++}(x^{+})dx^{+}dx^{+}
+ 2g_{+-}dx^{+}dx^{-}.\nonumber\end{equation} We are used Poisson
brackets (\ref{eq7}) for the space coordinates $x^{a}=
\{x^{i},x^{+},x^{-}\},\,i=1,2...D-2.$ We introduced the light-cone
gauge as two first kind constraints:
\[F_{1}(\sigma)=x'^{+}\approx 0,\,\,F_{2}(\sigma)=p'_{-}\approx 0,\]
and we imposed them on the equations of motion and on the Jacobi
identities. The Jacobi identities are reduced to the simple
equation
\[\frac{\partial \omega^{ab}}{\partial x^{+}}\omega^{+c}-\frac{\partial
\omega^{ac}}{\partial x^{+}}\omega^{+b}=0.\] 
We obtained following
result from this equation and additional condition (\ref{eq8}): there is 
constant background gravity field only
for the non-degenerate metric.
\section {Constant background fields ($g_{ab}=const.$).}
 In this section we are supplemented the bi-Hamiltonity condition
(\ref{eq6}) by the mirror transformations of the integrals
of motion.
\[\dot x^{a}=\{x^{a},\int\limits_{0}^{\pi}h_{1}d{\sigma}'\}_{1}=
\{x^{a},\int\limits_{0}^{\pi}{\pm h_{2}}d{\sigma}'\}_{\pm 2}.\]
 The dual PB are
\[\{x^{a}(\sigma ),x^{b}({\sigma}')\}_{\pm 2}=
\pm g^{ab}\nu ({\sigma}' -\sigma ),\,\, \{x^{a}(\sigma),p_{b}
({\sigma}' )\}_{\pm 2}=0,\]
\[\{p_{a}(\sigma ),p_{b}({\sigma}' )\}_{\pm 2}
=\mp g_{ab}\frac{\partial}{\partial {\sigma}' }\delta ({\sigma}'
-\sigma ).\]
 The dual dynamical system
\[\dot x^{a}=\{x^{a}, \pm H_{2}\}_{1}=\{x^{a}, H_{1}\}_{\pm 2}.\]
is the left(right) chiral string
\[\dot x^{a}=\pm x^{'a},\,\, \dot p_{a}=\pm p_{a}^{'}.\]
 Another way to obtain the dual brackets is the imposition of the second
kind constraints on the initial dynamical system, by such manner,
that $F_{i}=F_{k}$ for $i\ne k, i, k = 1,2,...$ on the constraints
surface $f(x^{a},p_{a})=0$.\\
{\bf Example 3}
The constraints $f^{(-)}_{a}(x,p)=p_{a}-g_{ab}x'^{b}\approx 0$ or
$f^{(+)}_{a}=p_{a}+g_{ab}x'^{b}\approx 0$ (do not simultaneously)
are the second kind constraints.  \[\{f^{(\pm)}_{a}(\sigma),
f^{(\pm)}_{b}({\sigma}')\}_{1}=C^{(\pm)}_{ab}(\sigma -{\sigma}')=
\pm 2g_{ab}\frac{\partial}{\partial {\sigma}'}\delta ({\sigma}'
-\sigma).\]
 The inverse matrix $(C^{(\pm)})^{-1}$ has following form
$C^{(\pm)ab}(\sigma -{\sigma}')=\pm\frac{1}{2}g^{ab}\nu({\sigma}'
-\sigma).$
 There is only one set
of the constraints, because consistency condition
\[\{f^{(\pm)}(\sigma),
H_{1}\}_{1}= f^{'(\pm)}(\sigma)\approx 0,\,\,...\,\,,
\{f^{(\pm)(n)}(\sigma), H_{1}\}_{1}= f^{(\pm)(n+1)}(\sigma)
\approx 0.\nonumber\] is not produce the new sets of constraints.
 By using the
standard definition of the Dirac bracket, we are obtained
following Dirac brackets for the phase space coordinates.
\[\{x^{a}(\sigma),
x^{b}({\sigma}')\}_{D}=\pm\frac{1}{2}g^{ab}\nu({\sigma}' -\sigma),
\{p_{a}(\sigma),p_{b}({\sigma}')\}_{D}=\mp\frac{1}{2}g_{ab}\frac{\partial}
{{\sigma}'}\delta({\sigma}'-\sigma),\]
\[\{x^{a}(\sigma),p_{b}({\sigma}')\}_{D}=\frac{1}{2}\delta^{a}_{b}
\delta({\sigma}' -\sigma).\] {equation}
 The equations of motion under the Hamiltonians $H_{1}=h_{1},H_{2}= h_{2}$ and
Dirac bracket \[\dot x^{a}=\{x^{a}, H_{1}\}_{D}=\{x^{a},
H_{2}\}_{D}=g^{ab}p_{b}=\pm x'^{a},\]
\[\dot p_{a}=\{p_{a},H_{1}\}_{D}=\{p_{a}, H_{2}\}_{D}=g_{ab}x'^{b}=
\pm p'_{a}.\] are coincide on the constraints surface. The dual
brackets $\{, \}_{\pm 2}$ are coincide with the Dirac brackets
also. The contraction of the algebra of the first kind constraints
means that the integrals of motion $H_{1}=H_{2}$ are coincide on
the constraints surface too.\\
{\bf Example 4} Constraints $f_{a}(\sigma)=p_{a}-h_{ac}x'^{c}(\sigma)$,where 
metric tensor of second fundamental form $h_{ac}=const., h_{ab}=h_{ba}$, 
$h_{ab}h^{bc}=\delta _{a}^{c}$ are the second kind constraints:
\[\{f_{a}(\sigma),f_{b}({\sigma}')\}=C_{ab}(\sigma - {\sigma}')=
2h_{ab}\frac{\partial}{\partial {\sigma}'}\delta ({\sigma}' - \sigma).\]
Inverse  matrix $C^{ab}(\sigma - {\sigma}')$ has form:
\[C^{ab}(\sigma - {\sigma}')=\frac{1}{2}h^{ab}\nu ({\sigma}' - \sigma).\]
Dirac bracket of arbitrary function $A(\sigma), B(\sigma)$ is
\[\{A(\sigma),B({\sigma}')\}_{D1}=\{A(\sigma),B({\sigma}')\}_{1} -\]
\[-\int\{A(\sigma),f_{a}({\sigma}'')\}_{1}C^{ab}({\sigma}''-{\sigma}''')
\{f_{b}({\sigma}'''),B({\sigma}')\}_{1}d{\sigma}''d{\sigma}'''.\]
Therefore, we obtained the only Dirac bracket
\[\{x^{a}(\sigma),x^{b}({\sigma}'\}_{D1}=-\frac{1}{2}h^{ab}
\nu (\sigma - {\sigma}')\]
on the surface $p_{a}-h_{ab}x^{'b}=0$.
The equations of motion under the Dirac bracket are
\begin{eqnarray}
\dot x^{a}(\sigma)=\int\{x^{a}(\sigma),h_{1}({\sigma}'\}d{\sigma}'\}_{D1}=
h_{b}^{a}x'^{b},\,\,\dot x^{a}(\sigma)=\int\{x^{a}(\sigma),
h_{2}({\sigma}'\}_{D1}d{\sigma}'=x'^{a}(\sigma),\nonumber\end{eqnarray}
Where $h^{a}_{b}=g^{ac}h_{cb}$ is Weingarten operator. The equation of motion
under the Hamiltonian $h_{1}$ 
\begin{equation}\label{eq16}\dot x^{a}=h^{a}_{b}x'^{b}\end{equation}
is the hydrodynamical type equation \cite{Ger:Mok2}. The equation 
(\ref{eq16}) for the
diagonal operator $h^{a}_{b}=\delta ^{a}_{b}h_{b}$ was considered as the 
Hamiltonian equation under the local bracket for the sphere embedded in a 
pseudo-Euclidean space $E^{N}$. 
It has the  following form in the sphere - conic coordinates $R^{a}$:  
\cite{Ger:Fer2,Ger:Pav}.
\[\dot R^{a}=(2R^{a}+\sum_{k=1}^{N-1} R^{k} - 
\sum_{K=1}^{N}h^{a})R'^{a}\]
The bi-Hamiltonity condition 
\[\dot x^{a}=\int\{x^{a}(\sigma),h^{1}({\sigma}')\}_{D1}d{\sigma}'=
\int\{x^{a}(\sigma),h_{2}({\sigma}')\}_{D2}d{\sigma}'\]
led to the following dual Dirac brackets:
\begin{eqnarray}\label{eq17}
\{x^{a}(\sigma),x^{b}({\sigma}')\}_{D2}=\frac{1}{2}g^{ab}
\nu ({\sigma}' - \sigma),\,\,
\{p_{a}(\sigma),p_{b}({\sigma}')\}_{D2}=-\frac{1}{2}g_{ab}\frac{\partial}
{\partial {\sigma}'}\delta({\sigma}' - \sigma),\nonumber \\
\{x^{a}(\sigma),p_{b}({\sigma}')\}_{D2}=\frac{1}{2}h^{a}_{b}\delta (\sigma -
{\sigma}').\end{eqnarray} 
\vskip 0.2cm
\section { Acknowledgments}

\end{document}